# A wavelength and polarization selective photon sieve for holographic applications

*Daniel Frese[1,#], Basudeb Sain[1,#], Hongqiang Zhou[2,#], Yongtian Wang[2],*

*Lingling Huang[2,\*], and Thomas Zentgraf[1,†]*

[1] Department of Physics, Paderborn University, Warburger Straße 100, 33098 Paderborn, Germany

[2] School of Optics and Photonics, Beijing Institute of Technology, 100081, Beijing, China

[#]These authors contributed equally to this work.

**Abstract:**

Optical metasurfaces are perfect candidates for the phase and amplitude modulation of light, featuring an excellent basis for holographic applications. In this work, we present a dual amplitude holographic scheme based on the photon sieve principle, which is then combined with a phase hologram by utilizing the Pancharatnam-Berry phase. We demonstrate that two types of apertures, rectangular and square shapes in a gold film filled with silicon nanoantennas are sufficient to create two amplitude holograms at two different wavelengths in the visible, multiplexed with an additional phase-only hologram. The nanoantennas are tailored to adjust the spectral transmittance of the apertures, enabling the wavelength sensitivity. The phase-only hologram is implemented by utilizing the anisotropic rectangular structure. Interestingly, such three holograms have quantitative mathematical correlations with each other. Thus, the flexibility of polarization and wavelength channels can be utilized with custom-tailored features to achieve such amplitude and phase holography simultaneously without sacrificing any space-bandwidth product. The present scheme has the potential to store different pieces of information which can be displayed

[\*] Email: huanglingling@bit.edu.cn

[†] Email:Thomas.zentgraf@uni-paderborn.de

separately by switching the wavelength or the polarization state of the reading light beam.

**Introduction:**

Optical metasurfaces provide outstanding opportunities in light shaping applications within nanometer dimensions. Carefully designed subwavelength scaled plasmonic and dielectric nanoscatterers within metasurfaces can enable the amplitude, phase and wavelength manipulation of the electromagnetic waves with high degrees of freedom[1-3]. In the past two decades, metasurfaces have been recognized as an excellent platform for versatile applications in polarizations optics[4-6], focusing elements[7-12], vortex beam generation[13-15], quantum optics[16-17], holography[18-22] and many more[23-24]. Due to the design flexibility for phase and amplitude modulation, metasurface holography has appeared as an important element for display applications[25-27]. With the opportunity to include manyfold information capability, several holographic multiplexing techniques based on metasurfaces have been developed. For instance, by using polarization multiplexing techniques, several phase masks can be encoded and the different pieces of information are reconstructed in different polarization channels, resulting in a single device with several information channels[28-29]. To include color operation, which is attractive for the human eye, metasurfaces can either be subdivided into RGB regions to generate colored images[30-31], or nonlinear material properties of the individual structures can be utilized to create different colored images[32-33].

Recently, metaholograms based on photon sieves have been demonstrated as an attractive possibility to create amplitude holograms[34-35]. Photon sieves are diffractive devices constisting of a metal film with nanoholes in a similar arrangement like the rings of Fresnel zone plates[36]. It was originally developed for applications in X-ray focussing[37], where it is shown that the size and distribution of the nanoholes affect the amplitude modulation of the transmitted light. Thanks to modern algorithms, the arrangements for a desired light modulation can be calculated and the application range expanded to vortex beams[38] and holograms[34-35]. However, the photon sieves demonstrated so far are generally polarization-independent and broadband. Furthermore, multiplexing techniques for additional phase modulations or wavelength selective amplitude modulation have not been applied to obtain switchable and multifunctional devices, which would be beneficial for holographic displays.

Here, we present a modified photon sieve hologram operating at two distinct wavelengths, green and red, combined with a geometric phase hologram. In contrast to a classical photon sieve with isotropic circular nanoholes, we select two different aperture types in a aperiodic arrangement and fill those with anisotropic silicon nanoantennas, which are selected based on their spectral behaviors. Using the two different aperture types, a rectangular and a square one, enables multiplexing of two amplitude-only holograms and one phase-only hologram into a single device. By filling the rectangle and square nanoholes with silicon nanoantennas, the spectral transmittance of the apertures can be tuned depending on the silicon nanoantenna dimensions. Thus, in contrast to broadband operation in the classical case of a photon sieve, we can implement custom-tailored features at selected wavelengths as well as selected polarization channels. In our case, we tune the two different aperture types through the nanoantennas' dimensions, resulting in two different wavelengths to encode two distinct amplitude holograms in the circular co-polarization state. In addition, by utilizing the concept of the Pancharatnam Berry (PB) phase for the rectangular (anisotropic) shape, the third (phase) hologram is encoded in the circular cross-polarization channel. Thus, we show that the classical broadband operation of the photon sieves can be extended by holographic multiplexing techniques, which can potentially be applied to holographic display applications, as well as optical data storage and security features.

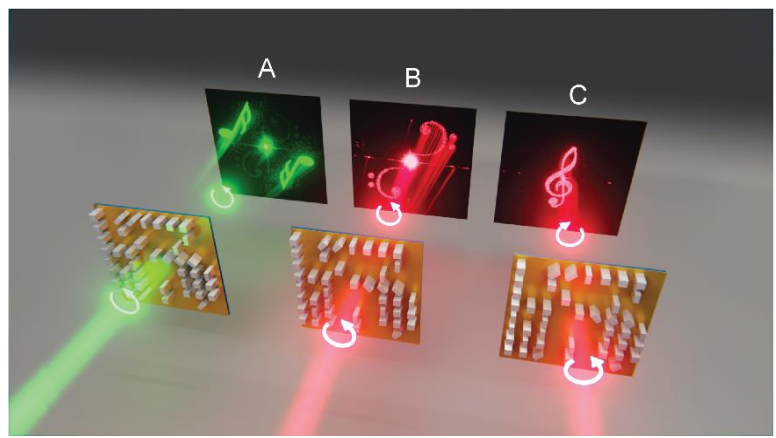

**Figure 1: Illustration of the modified photon sieve metahologram working principle.** The modified photon sieve holograms are based on the aperiodic arrangements of rectangular and square-shaped gold apertures filled with silicon nanoantennas to form a metasurface

hologram. By tuning the transmittance of the individual structures, as well as the orientation angle of the rectangular shapes, a double amplitude hologram (*A* and *B*) combined with a phase-only hologram (*C*) is realized.

**Metasurface Design and Working Principle**

In this work, we present a modified aperiodic photon sieve arrangement, where the aperture shapes are chosen as square and rectangular structures, in contrast to nanoholes in the case of classical photon sieves. As illustrated in Figure 1, the transmission properties of the different aperture types, filled with the silicon nanoantennas, can be tuned to achieve particular transmission properties to encode two different amplitude holograms *A* (music notes) and *B* (bass clefs) at wavelengths $\lambda_1$ and $\lambda_2$, respectively, and one phase-only hologram *C* (violin clef) based on the Pancharatnam-Berry phase effect.

A single binary amplitude hologram contains two bits of information, which are 0 or 1 for the transmission modulation. However, such two bits of information can be realized by using different wavelengths, polarizations, spatial overlapping, et al. as information channels. Generally, there are four possible combinations for a metasurface pixel to contribute to two independent binary amplitude holograms, that is, (0, 0), (0, 1), (1, 0), and (1, 1). In our case, the wavelengths ($\lambda_1$, $\lambda_2$) form the information channel for the amplitude holograms in the circular co-polarization state. We develop a quantitative correlated relationship between the two amplitude holograms to reduce the number of combinations. Related to our designed metasurface hologram, the bare gold film of the photon sieve does not transmit light and is represented by (0, 0). The square structure transmits to both amplitude holograms, resulting in (1, 1), while the rectangle structure only transmits light to one of the amplitude holograms and results in (1, 0). The remaining combination of (0, 1) can be canceled by building the mathematical relation of $B \supset A$, meaning that the square structures form the hologram *B* when illuminated with monochromatic light of wavelength $\lambda_2$, but also contribute as a subset to hologram *A* at $\lambda_1$. Note, that in both cases the gold film (0, 0) is mandatory necessary to form the amplitude holograms. By utilizing the cross-polarization conversion of the anisotropic rectangular structure, we can encode a third phase-only hologram in the circular cross-polarization channel based on the Pancharatnam-Berry phase. By carefully designing the dimensions of the rectangular silicon antenna, one can realize that the nanoantenna contributes to *A* at $\lambda_1$ in co-polarization and form another arbitrary hologram *C* at $\lambda_2$ in the cross-polarization channel. The overall relationship of the holograms can be expressed as $B \supset A$, $A = B \cup C$, while $B \cap C = 0$.

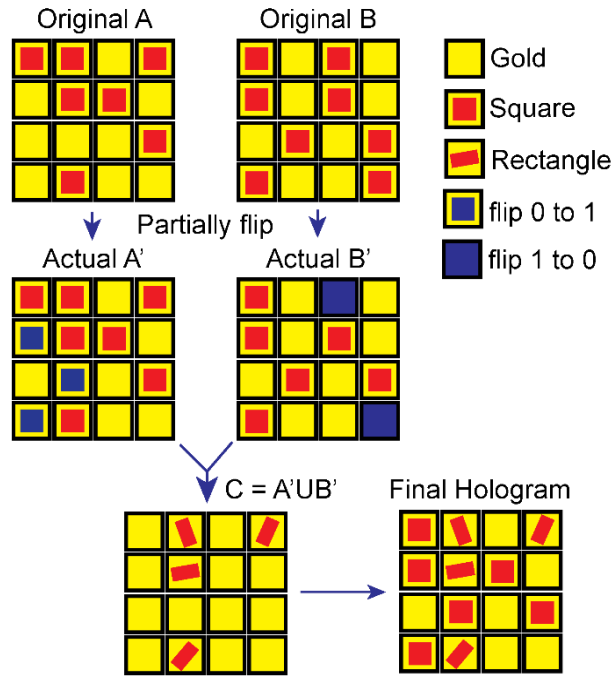

**Figure 2: Schematic diagram of the hologram algorithm.** After the two holograms Original A and Original B are formed, random flips of transparent to transmitting pixels and vice versa are applied to create Actual A' and Actual B' under comparison to the original holograms. The amplitude multiplexing is achieved by replacing all squares in A' that are not included in B' through rectangulars, forming a wavelength sensitivity based on the individual structures' spectral properties. An arbitrary phase-only hologram is finally formed in the cross-polarization channel by rotating the rectangular nanoantennas.

To achieve the above scheme, we develop a novel holographic algorithm based on the traditional Gerchberg-Saxton algorithm[39]. As shown in Fig. 2, first, two independent binary amplitude holograms (Original *A* and Original *B*) are generated with random binary amplitude noise. Each pixel of the holograms can either transmit light, represented by the red square structure or block the light, represented by the gold film in yellow. Second, we partially flip the pixels (from tranmissive to not transmissive or vice versa) of hologram *A* and *B*, resulting in the actual holograms *A*' and *B*', and compare those with the original ones in each iteration, ensuring that the aberration from the original to the actual holograms are sufficiently small. Thus, due to the robustness of holograms, one can still reconstruct the holograms *A* and *B* from *A*' and *B*' by flipping a decent amount of pixels in each iteration. Further, we compare *A*' and *B*' pixel by pixel and form the basis of hologram *C* by replacing all squares in *A*' by the rectangular structures, that are not included in *B'* (*C = A'* ∩ *B'*). Thus, we end up with the final metasurface hologram that transmits light through the square and rectangular structures at $\lambda_1$ (forming *A*) and only transmit light through the square structure at $\lambda_2$ (forming *B*) in the co-polarization channel. In addition, we can implement a third phase-only hologram *C* in the cross-polarization channel based on the Pancharatnam-Berry phase by rotating the rectangular structures, which does not influence the transmittance of the co-polarization channel. Finally, we end up with a 1200 by 1200 pixels hologram, where 47.91 % of the pixels are filled with gold, 31.24 % are filled with the square shaped structures and 20.85 % are filled with the rectangular structures.

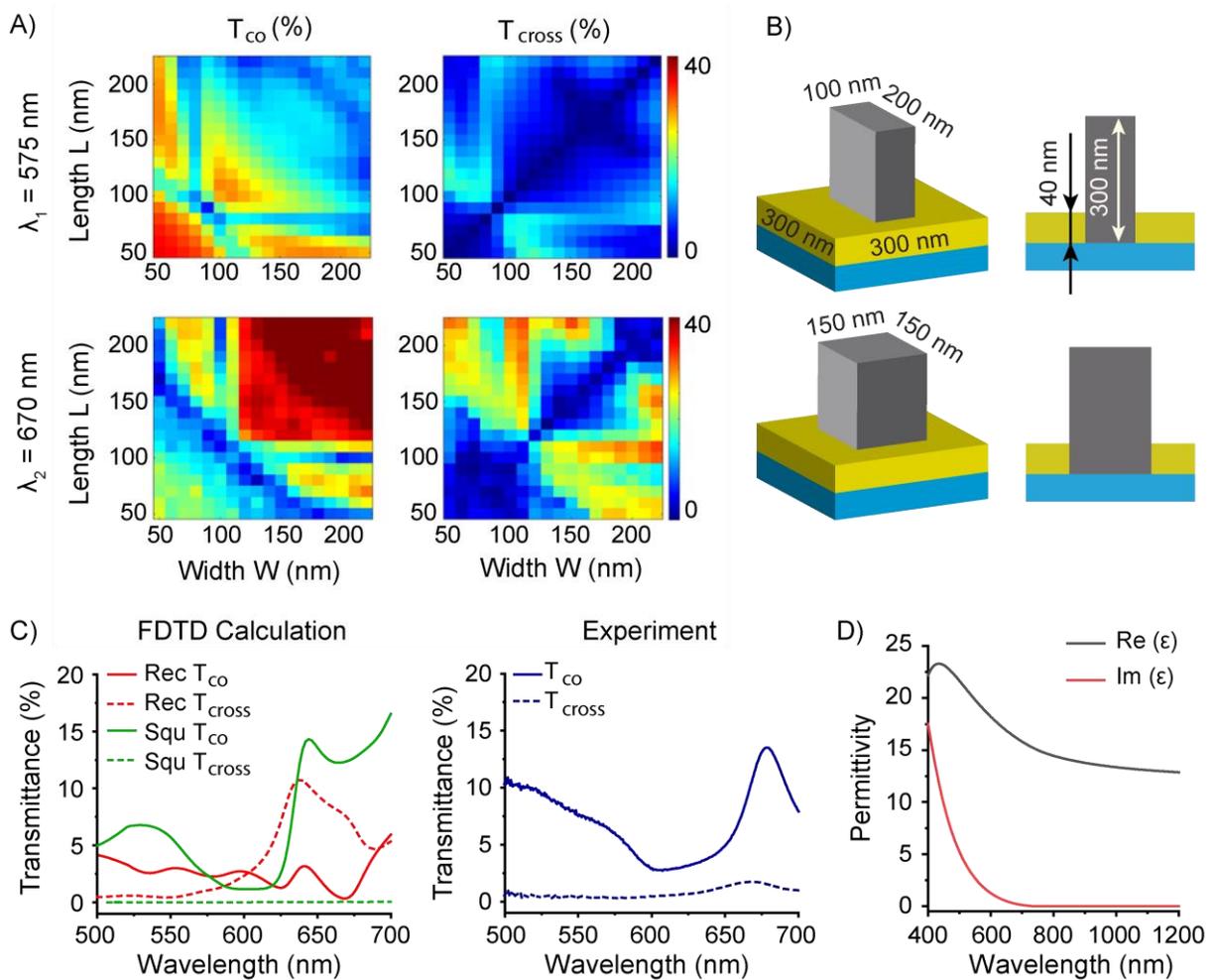

**Figure 3: Simulated and experimentally measured transmittance.** A) Parameter sweep to obtain the optimized silicon antenna dimensions L and W for a set height of 300 nm. The transmittance is calculated in the co- and cross-polarization channels at the design wavelengths $\lambda_1$ and $\lambda_2$. B) Selected antenna dimensions for the rectangular and square nanoantennas illustrated in 3D and the corresponding 2D cross-sections. C) Left: Calculated transmittance of the selected silicon antennas (Rec: 100 nm x 200 nm x 300 nm; Squ: 150 nm x 150 nm x 300 nm) including the gold film in the co- and cross-polarization channel with finite-

difference time-domain method (FDTD). Right: Measured transmittance of the fabricated modified photon sieve in the aperiodic arrangement for both polarization channels. D) Measured permittivity of amorphous silicon used for the simulations.

To design the sample, we started with the investigation of the single nanoantennas, including the gold film in a square lattice and calculated the transmittance using rigorous coupled-wave analysis (RCWA). The length L and width W of the nanoantennas are swept from 50 nm to 220 nm at the two working wavelengths of $\lambda_1$ = 575 nm and $\lambda_2$ = 670 nm, as shown in Fig. 3A. The square silicon nanoantennas are designed to be transparent pixels and transmissive at $\lambda_1$ and $\lambda_2$. The rectangular nanoantenna is optimized based on its wavelength and polarization sensitivity. In the co-polarization channel, such anisotropic nanoantenna is designed to transmit light at the wavelength $\lambda_1$, while in cross-polarization, it is designed to transmit light at $\lambda_2$. We choose the dimensions L = 150 nm and W = 150 nm for the square structure and L = 200 nm and W = 100 nm for the rectangular structure (for a sketch of the structures see Fig. 3B). The calculated transmittance of the individual structures in a square lattice arrangement is shown in Fig. 3C, left. In the co-polarization channel, the solid red and green line intersect at 575 nm with a transmittance of 2 %, while at 670 nm the square structure shows a significantly higher transmittance of 13 % compared to the rectangular structure (1 %). Compared to the experimentally measured transmittance of the fabricated sample (Fig. 3C, right), the sample transmits more light as expected, even though the filling ratio of the rectangular and square structures is less compared to the ideal (calculated) case. The transmittance at 575 nm and 678 nm (local maximum) are 6 % and 14 %, respectively, which are equivalent to the holograms' efficiencies in the co-polarization channel. The overall deviations from the simulation may result from the fabrication tolerances, as well as the different arrangements in the simulation compared to the experiment. In the experimentally realized structure, the interactions between the individual structures might be strongly reduced due to their aperiodic arrangement and the pixels without nanoantennas. Note, that we use the permittivity of amorphous silicon in our simulations, measured based on ellipsometry, plotted in Fig. 3D.

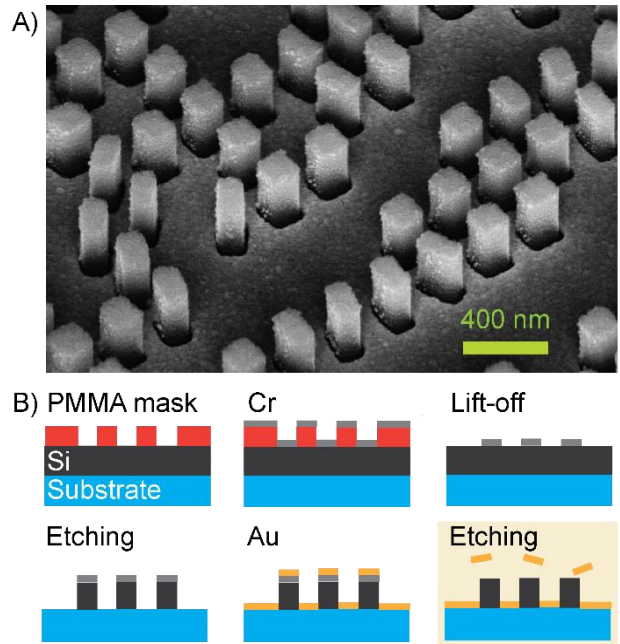

**Figure 4: Scanning electron microscope image (SEM) of the fabricated sample and flow chart diagrams of the fabrication process.** A) SEM image of the fabricated modified photon sieve sample, showing the silicon nanostructures standing within a homogeneous gold film. B) Flow chart of the fabrication processes: After transferring the desired pattern to the photoresist (PMMA) by e-beam lithography, the hard etching mask was formed by the subsequent deposition of chromium and a lift-off process. Using an inductively coupled reactive-ion etching (ICP-RIE) process, the desired patterns were transferred to the silicon with a chromium mask on top of the individual silicon nanopillars. Next, a 40 nm gold layer was deposited. The gold cups on top of the silicon nanopillars were removed by selectively etching away the sacrificial chromium layer.

**Results:**

To reconstruct the three different encoded Fourier holograms, we use a 4f-setup to image the k-space of the metasurface (Fig. 5A). For that, we use a supercontinuum light source with a monochromator as a tunable wavelength source in the visible spectrum with a bandwidth of 5 nm. The laser light is converted into a circular polarization state using a linear polarizer followed by a 45° rotated quarter-wave plate. After the light is focused on the metasurface by a lens with a 500 mm focal length, it is collected by a microscope objective with 20x magnification and NA=0.7. The back focal plane of the objective is projected on a monochrome camera by using two 100 mm lenses as illustrated. A second polarizer unit, placed behind the microscope objective is used to select either the co- or the cross-polarization state.

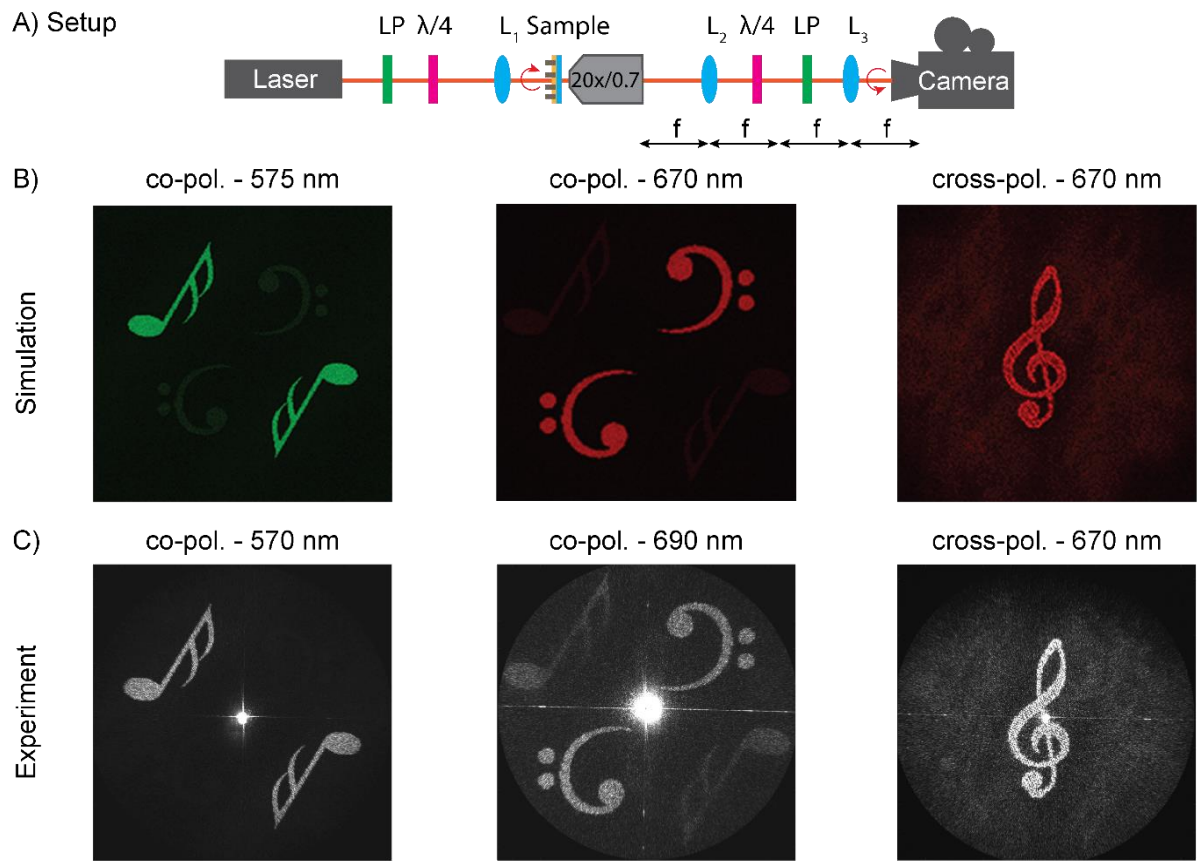

**Figure 5: Measured holograms and their target images.** A) The sample is illuminated with laser light. The wavelength can be tuned and each set wavelength has a bandwidth of 5 nm. The input polarization state is circular, while the output state can be changed from the co- to the cross-circular polarization state. The encoded Fourier-holograms are imaged to the monochrome camera using a 4f-setup. B) Simulated holograms. The music notes and bass clefs are the dominant target image at 575 nm and 670 nm in the circular co-polarization state. In the cross-polarization state, at 670 nm, the violin clef appears based on the Pancharatnam-Berry phase. C) Experimentally measured holograms.

The simulated reconstructed images are shown in Figure 5B. At $\lambda_1 = 575$ nm, we encoded two music notes (hologram *A*), off-axis, in the k-space in the co-polarization state, while at $\lambda_2 = 670$ nm, bass clefs (hologram *B*) are encoded in the co-polarization state. In addition, at 670 nm, the rectangular structure is arranged to carry the geometric phase hologram of a violin clef (hologram *C*), which can be reconstructed in the cross-polarization state. To calculate the hologram reconstruction, we consider the transmittance of the FDTD simulation at the desired wavelengths and polarizations. The measured holograms in the co- and cross-circular polarization state are shown in Fig. 5 C. At 570 nm, the music notes are imaged homogeneously and with very low background. The 0$^{th}$-order sport in the middle of the image originates from unconverted light which passes the metasurfaces. At 690 nm, the intensity ratio between the bass clefs and the music notes is the highest. The holographic feature is redshifted by 20 nm compared to the design at 670 nm. Note, that the sub-images cannot be ignored completely in the measured holograms, since we tolerate a slight amount of cross-talk in the design by iteratively flipping partial pixels from 0 to 1 and vice versa, as explained above. Further, the music notes (*A*) are a subset of the bass clefs (*B*), as well as the transmittance of the rectangular structures is not equal to zero for 690 nm. Thus, from a theoretical as well as from the physical point of view, the unwanted sub-images cannot be ignored completely. However, in both cases of the amplitude holograms, the main holograms are dominant and coincide with the calculated target images. In the case of the music notes at 570 nm, the sub-images are barely visible. In the cross-polarization state, the violin clef (hologram *C*) can be reconstructed with good agreement to its calculated target image. The experimentally measured conversion efficiency is 2 % at 570 nm. The small $0^{th}$-order spot can result from imperfect polarization conversion based on the sample fabrication, as well as from unideal components used in the optical setup.

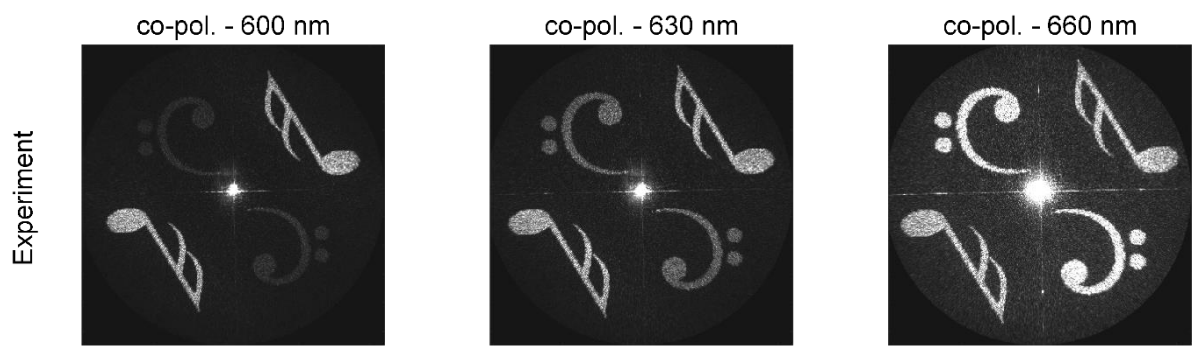

**Figure 6: Measured holograms at wavelengths deviating from the design wavlengths.** Through the different sperctaral behavior of the rectangular and square structure, the hologram behavior changes by changing the input wavelength other than the design wavelength, as can be seen in the changing brightness of the partial holograms.

To illustrate the behaivior of the photon sieve hologram at wavelengths different from the design wavelength, we measured the holograms at 600 nm, 630 nm and 660 nm in the co-polarization channel (Fig. 6). We observe a change in the brightness of both sub-holograms, originating from the spectral behavior of the two different nanoantenna types involved. As expected, if the ratio of transmitted light through the rectangular and square structure deviates from the original design, both holograms are projected on the camera depending on the transmittance of the individual structures. This behavior emphasizes the necessity of the silicon nanoantennas within the nanoapertures, enabling the design of the large contrast between the transmittance of the rectangular and square shaped nanostructures at 690 nm, where the transmittance of the rectangular structures drops close to zero. Note, that also other materials than silicon can be used to obtain a similar effect. However, the overall agreement of the measured holograms with the target images illustrates how amplitude and phase holograms can be implemented in a single device to enlarge the information density of holograms, as well as switching images by changing the illuminating light beam properties, like wavelength or polarization.

**Conclusion:**

In conclusion, we realized a dual amplitude holographic metasurface based on the photon sieve principle with two distinct wavelengths of reconstruction, multiplexed with a phase-only hologram based on the Pancharatnam-Berry phase. Therefore, we utilized the transmission features of two sets of silicon nano-antennas placed in a gold film. While the gold film blocks the light between the nano-apertures as in the case of a classical photon sieve, the two different types of silicon nanostructures, with rectangular and a square cross-section, transmit the desired light. We show, that the transmission properties of the photon sieve apertures can be tuned by the silicon antennas to achieve custom-tailored transmission properties. Despite the relatively low efficiency that can be derived from the transmittance, the dual amplitude holograms can be reconstructed with outstanding image quality. Thereby, the generally polarization-independent and broadband operation of classical photon sieves can be expanded by holographic multiplexing techniques to provide wavelength-selective operation at multiple wavelengths and, by utilizing the geometries of the apertures, integrate polarization-dependent information channels. Thus, these devices can be switched in dependence of wavelength and polarization and become more flexible for applications in switchable holographic displays. Furthermore, the information density of photon sieves can be enlarged by multiplexing technologies, since the different structures can be designed to operate in several information channels simultaneously.

**Method**

The fabricated structure is shown in the scanning electron microscopy (SEM) image in Fig. 4A. Under 45° view, one can see the aperiodic arrangements and that the silicon nanoantennas are standing out of the gold film. To fabricate the sample, we use a standard process based on electron-beam lithography (Fig. 4B). First, a chromium mask is formed and the silicon nanostructures are etched using an inductively coupled reactive ion etching (ICP-RIE) process. After etching, the 40 nm gold layer is deposited. To remove the unwanted metals from the top of the silicon nanostructures, we utilize the chromium as a sacrificial layer and put the sample in commercial chromium etch solution. Thus, the 40 nm chromium film is etched and the gold cups sitting on top of the chromium layer get detached from the nanostructures.

**Acknowledgment**

This project received funding from the Deutsche Forschungsgemeinschaft (DFG, German Research Foundation) – SFB-Geschäftszeichen TRR142/2-2020 – Projektnummer 231447078 – Teilprojekt A08. The authors acknowledge the funding provided by the National Key R&D Program of China (No. 2017YFB1002900) and the European Research Council (ERC) under the European Union's Horizon 2020 research and innovation program (grant agreement No. 724306). We also acknowledge the NSFC-DFG joint program (DFG No. ZE953/11-1, NSFC No. 61861136010). L.H. acknowledges the support from the Beijing Outstanding Young Scientist Program (BJJWZYJH01201910007022) and the National Natural Science Foundation of China (No. 61775019, No. 92050117) program.